\documentstyle[psfig]{EuroPhys}
\input EuroMacr
\input epsf

 \def\gappeq{\mathrel{\rlap {\raise.5ex\hbox{$>$}}
{\lower.5ex\hbox{$\sim$}}}}
\def\lappeq{\mathrel{\rlap{\raise.5ex\hbox{$<$}}
{\lower.5ex\hbox{$\sim$}}}}

\begin{document}
\def\pa{\parallel}
\def\pe{\bot}
\def\xx{{\bf{x}}}     
\shorttitle{Growth and Scaling in Anisotropic Spinodal Decomposition}
\title{ \large \bf GROWTH AND SCALING IN ANISOTROPIC SPINODAL DECOMPOSITION}
\author{ P. I. Hurtado$^1$, J. Marro$^1$, E. V. Albano$^2$}     
\institute{
$^1$ Instituto {\em Carlos I} de F\'{\i}sica Te\'{o}rica y Computacional,\\
and Departamento de Electromagnetismo y F\'{\i}sica de la Materia, \\
Universidad de Granada, 18071-Granada, Spain.\\
$^2$ Instituto de Investigaciones Fisicoqu\'{\i}micas Te\'{o}ricas y
Aplicadas, UNLP,\\
CONICET, Sucursal 4, Casilla de Correo 16, (1900) La Plata, Argentina.
}
\rec{}{}
\pacs{
\Pacs{05}{20.Dd}{Kinetic theory}
\Pacs{61}{20.Ja}{Computer simulation of liquid structure}
\Pacs{64}{60.Qb}{Nucleation}
}
\maketitle

\begin{abstract}
We studied phase separation in a particle interacting system under a large
drive along $x$. We here identify the basic growth mechanisms, and
demonstrate time self-similarity, finite-size scaling, as well as other
interesting features of both the structure factor and the scaling
function. We also show that, at late $t$ in two dimensions, there is a unique
$t-$\textit{dependent} length increasing $\ell _{y}\left( t\right) \sim t^{1/3}$ for
macroscopic systems. Our results, which follow as a direct consequence of the
underlying anisotropy, may characterize a class of nonequilibrium situations.
\end{abstract}
\vspace{4pt}

%%%%%%%%%%%%%%%%%%%%%%%%%%%%%%%%%%%%%%%%%%%%%%%%%%%%%%%%%%%%%%%%%%%%%%%%%%%%%%%%%%%

Many binary mixtures, e.g. the alloy Al-Zn, which are homogeneous at high
temperature, develop coarsening macroscopic grains after a quench into the
miscibility gap. The details of nucleation and spinodal decomposition as the
system evolves towards coexistence of the two new phases determines various
properties; e.g. hardness and resistivity of the alloy depend on how phase
separation kinetics competes with the progress of solidification from the melt.

The involved essential physics is rather well understood, partly due to
computer simulation of lattice gases.\cite{rev,prl00} An interesting task is
now understanding more general situations, e.g. lack of isotropy, which
bears great technological importance. Mixtures under a shear flow attracted
considerable attention as a possible scenario.\cite{critic}-\cite{corberi}
We studied the kinetics of the driven lattice gas (DLG) \cite{katz}
by extensive Monte Carlo (MC) simulation. This is appealing on several
grounds. Firstly, the DLG is the most successful, microscopic metaphor for
(nonequilibrium) anisotropic phenomena, and its time relaxation remains
intriguing.\cite{md}-\cite{albano} Furthermore, the common underlying
anisotropy might induce essential macroscopic similarities between the DLG
and the sheared lattice gas,\cite{chino} as recent results on critical
behavior seem to suggest.\cite{prl01,albano} In any case, the DLG is not
affected by hydrodynamic effects, which makes physics simpler.\cite{rev}

The DLG is a $d-$dimensional lattice gas (or, alternatively, binary alloy)
at temperature $T$ in which nearest-neighbor (NN) particle/hole exchanges
are favored along one of the principal lattice directions, say $\vec{x}$.\cite{md} 
There are variables $n_{i}=1$ (\textit{particle}) or $0$ 
(\textit{hole}) at each site $i=1,\ldots ,N$, a NN interaction according to 
$H=-4\sum_{NN}n_{i}n_{j}$, toroidal boundary conditions, and a
transition rate given (e.g.) by a $-$biased$-$ Metropolis algorithm, $\omega
({\mathbf n}\rightarrow {\mathbf n}^{\ast })=\min {\{1,\exp [-(\Delta
H+E\delta )/T]\},}$ which conserves density $\rho =N^{-1}\sum_{i}n_{i}$.
Here, $\mathbf{n}^{\ast }$ represents configuration $\mathbf{n=}\{n_{i}\}$
after jumping of a particle to a NN hole, $E\vec{x}$ may be interpreted as
an (electric) field driving (charged) particles, $\Delta H=H({\mathbf n}%
^{\ast })-H({\mathbf n})$, and $\delta =(\mp 1,0)$ for jumps along $\pm \vec{x%
}$ or along any of the transverse directions, say $\vec{y}$, respectively.

The DLG was described as modelling fast ionic conduction, surface growth,
traffic flow, etc.\cite{md} A common feature of all these situations is
anisotropy, and that steady states are out of equilibrium; both are
essential features of the DLG induced by the rate $\omega $. In general, 
$\omega $ violates detailed balance. This symmetry holds only for $E=0;$ the
DLG then reduces to the familiar lattice gas with a unique (equilibrium)
steady state independent of $\omega (\mathbf{n}\rightarrow \mathbf{n}^{\ast
})$. For any, even small $E$ the steady state depends on $\omega $, and a
different, qualitatively new behavior emerges.\cite{prl01} As $E$ is
increased, one eventually reaches saturation (particles cannot jump
backwards, $-\vec{x})$, which is formally denoted as $``E=\infty "$.

For $d=2$, $\rho =\frac{1}{2}$ and $E=\infty $ (the only case to which we refer
here for simplicity --also because this is a most interesting case \cite{md}), 
a critical point occurs at $T=T_{C}^{\infty }\simeq
1.4T_{C}\left( E=0\right) $. Steady states below $T_{C}^{\infty }$ do not
correspond to coexistence of two thermodynamic (equilibrium) phases as for $%
E=0$. Instead, stable ordered states consist of one single stripe,
corresponding to the \textit{liquid}, rich-particle phase, and \textit{gas},
poor-particle phase filling the remainder of the system. The interface is
linear and flat, except for microscopic roughness (which slightly increases with
decreasing $E$). One measures a net current of particles along $%
\vec{x}$ (its intensity increasing with $T$, and changing slope at $%
T_{C}^{\infty })$.\cite{md}

Our simulations (fig.1) proceed by means of $\omega $ from a random state,
until one or sometimes a few stripes are obtained. The code includes a list
of $\lambda (t)$ particle-hole NN pairs from where the next move is drawn. Time
is then increased by $\Delta t=\lambda (t)^{-1}$, so that its unit or \textit{%
MC step} involves a visit to all sites on the average. Most evolutions are
at $T=0.8T_{C}(E=0)\simeq 0.6T_{C}^{\infty }$ for which clustering is rather
compact and, in practice, one can observe the full process of relaxation; we
are assuming that, as observed for equilibrium states,\cite{rev}-\cite{prl00}
time evolution has the same properties in a wide region of the miscibility
gap (see argument below). The lattice is rectangular, $L_{x}\times L_{y},$
with sides ranging from 64 to 256 (and, exceptionally, 512).

Starting from complete disorder, there is a very short regime in which small
grains form (fig. 1a). Typical grains are anisotropic, stretched along $\vec{%
x}$. One then observes a rapid coarsening to form macroscopic strings,
fig.1b. We skip details concerning such nucleation and early phase
separation.\cite{fnote2} We wish to notice, however, that sheared fluids
seem to depict a similar behavior. That is, initial formation of small
anisotropic clusters (inducing a larger growth rate along the flow than in
the other directions), and stringlike domains extending macroscopically have
been reported in fluids.\cite{critic,corberi} 

After the initial (anisotropic) nucleation regime, strings coarsen further
into well defined, relatively narrow stripes; fig.1c. The resulting
multistripe states are not stable. They are partially segregated, and tend
to relax towards the true stable state of one stripe. This may take a long
macroscopic (say, \textit{infinite}) time; in fact, the mean relaxation time
increases with system size (see also \cite{mukamel}). As in the equilibrium
case,\cite{rev} this is a consequence of the conservation of $\rho $, which
makes the interfaces to depend on each other. It is true that certain
individual runs sometimes block for a long time in a state with a few
stripes; however, it does not seem to correspond to the average behavior.
Typically, the number of stripes monotonically decreases with time, and the
whole relaxation can easily be observed in computer simulations if one waits
long enough (see caption for fig. 1).

We shall assume that further coarsening occurs by monomer diffusion. \textit{%
Liquid} stripes thus effectively diffuse and, eventually, collide and
coalesce with one neighbor. This implies evaporation of a \textit{gas}
stripe. Therefore, given the particle/hole symmetry, our assumption is
equivalent $-$though it allows for a more detailed description below$-$ to
assuming coarsening by stripe evaporation as in \cite{mukamel}.

\begin{figure}
\centerline{
\epsfxsize=4.5in
\epsfbox{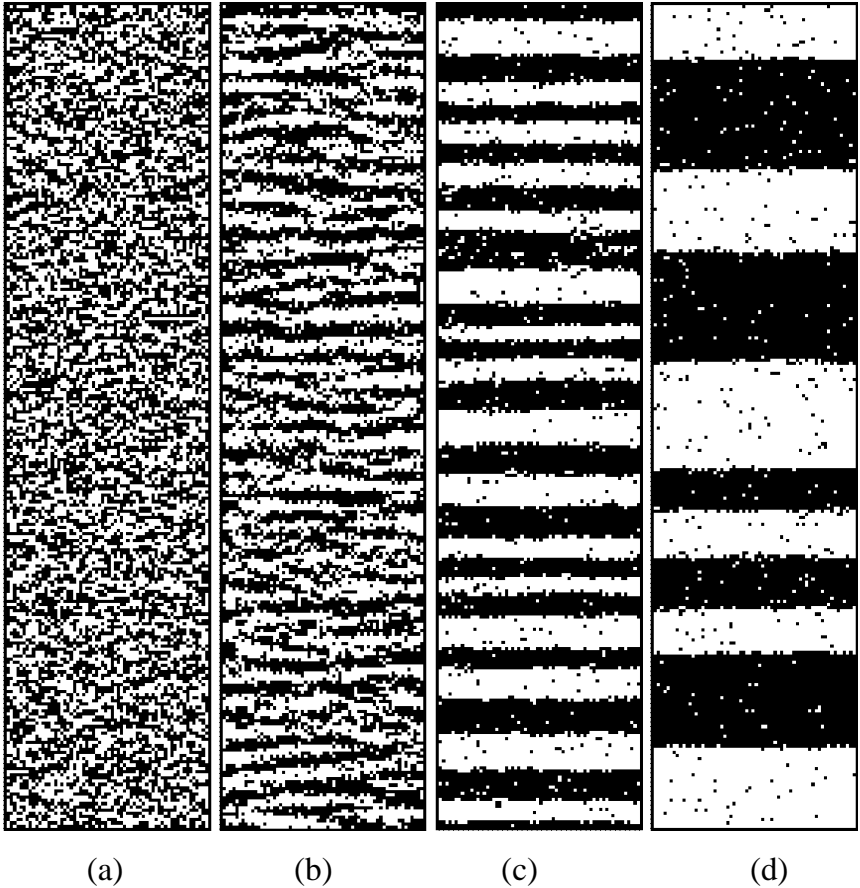}
}
\caption{A series of snapshots depicting growth at $T\simeq 0.6T_{C}^{\infty }$ 
for $L_{y}\times L_{x}=256\times 64$ and $t=1$ (a), 50
(b), 10$^{4}$ (c) and 2$\times $10$^{6}$ (d) MC steps. In this particular
run, two stripes were observed just before $10^{7}$ MC steps while a single
stripe was only reached after 10$^{8}$ MC steps.
}
\vspace{-0.5cm}
\label{figura1}
\end{figure}                                        

Let us evaluate sort of \textit{mobility} $\mathcal{D}_{\ell}$ due to stripe
diffusion via monomer events. Here $\mathcal{D}_{\ell}$ is the mean squared 
displacement of the stripe's center of mass per unit time.
Consider a compact stripe of mean width $\ell
(t)$ consisting of $j=1,\ldots ,M$ particles at $y_{j}(t)$. Its center of
mass is at $Y_{cm}(t)=M^{-1}\sum_{j}y_{j}(t)$. Then 
${\mathcal D}_{\ell}=N_{me} \langle (\Delta Y_{cm})^{2} \rangle $, 
with $N_{me}$ the frequency of events and $\langle (\Delta Y_{cm})^{2}\rangle $ 
the mean squared displacement associated with
one of them.\cite{binder} It ensues $\mathcal{D}_{\ell }$ as the result of two competing
processes: \newline
\noindent (A) A surface traps a monomer evaporated from the same interface.\cite
{gambler} Then $N_{me,A}=\nu \sum_{j}^{\prime }\exp (-2\beta \Delta
_{j})$ where $\nu $ is a frequency, the sum is over the surface particles,
$\beta$ is the inverse temperature 
and $\Delta _{j}$ is the number of broken bonds. For flat linear interfaces, 
$N_{me,A}\approx 4\nu L_{x}\exp (-2\beta \bar{\Delta})$ where $\bar{%
\Delta}$ is the mean of $\Delta _{j}$ (we here multiplied by 2 to account
for evaporation of surface holes reaching the surface again from the
interior). One further has $\Delta Y_{cm}=M^{-1}\delta y$, where 
$\delta y$ is the net displacement, and (for compact enough stripes) $%
M\approx L_{x}\times \ell (t),$ so that $\langle (\Delta Y_{cm%
})^{2}\rangle =\langle \delta y^{2}\rangle \ L_{x}^{-2}\ell ^{-2}$.
\newline
\ (B) A hole jumps within the stripe, inducing $\Delta Y_{cm}=1/M$ or
0 depending on the jump direction. One has $N_{me,B}=2\nu \rho
_{h}(T)L_{x}\ell p_{h}(T),$ where $\rho _{h}L_{x}\ell $ is the number of
holes in the stripe and $p_{h}$ is a jumping probability. At low $T$ (small
hole density $\rho _{h}$), holes are isolated from each other, so that 
$p_{h}\approx 1$. It ensues ${\mathcal D}_{\ell }^{(B)}\sim 2\nu \rho
_{h}L_{x}^{-1}\ell ^{-1}.$

As far as stripes are a distance $\ell $ apart from each other,
they take a time $\tau _{\ell }=\ell ^{2}/{\mathcal D}_{\ell }$ to meet,
which increases width by $\ell$. Then $d\ell /dt\sim {\mathcal D}_{\ell
}\ell ^{-1}$ and, assuming that processes A and B are independent, 

\begin{displaymath}
d\ell /dt\sim L_{x}^{-1}(\alpha _{A}\ell ^{-3}+\alpha
_{B}\ell
^{-2})
\end{displaymath}
for low $T$ and large $E;$ here, $\alpha _{A}=4\nu \langle \delta
y^{2}\rangle e^{-2\beta \bar{\Delta}}$ and $\alpha _{B}=2\nu \rho
_{h}$.

\begin{figure}
\centerline{
\epsfxsize=4.0in
\epsfbox{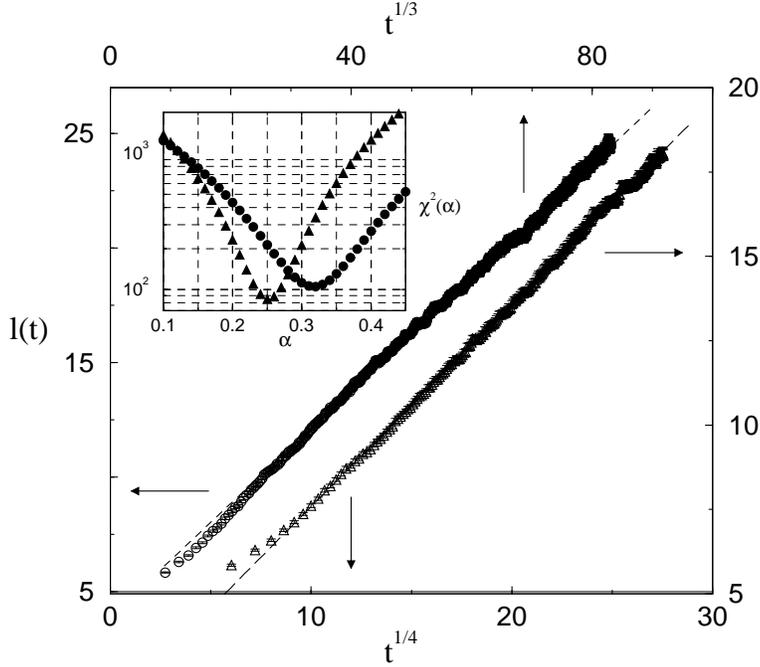}
}
\caption{The typical length $\ell \left(t\right) $ versus
 $t^{\phi _y }$ for a lattice of size 64$\times $64 ($\Delta $),
with $\phi _y =\frac{1}{4},$ and 256$\times 64$ (O), with $\phi _y =\frac{1}{3}.$
(Note that each set of data refers to different axis.) The inset shows 
Pearson Chi Square function, $\chi^{2}\left( \phi _y \right) $, for varying fits.
}
\vspace{-0.5cm}
\label{figura2}
\end{figure}                                        

This has some interesting implications. At late times, $\ell (t)\sim \alpha
t^{1/3}+\alpha _{A}/2\alpha _{B}$ with $\alpha ^{3}=3\alpha _{%
B}L_{x}^{-1}$, i.e., hole diffusion (mechanism B) is dominant.
Evaporation/condensation (mechanism A) results in $t^{1/4}$ behavior; this
is predicted to matter earlier. The crossover between the two
mechanisms is for $t\sim \tau _{cross}=(4\alpha _{A})^{3}(3\alpha
_{B})^{-4}L_{x}$, i.e., a macroscopic, observable time. If we define the
time at which a single stripe forms, $\ell (\tau _{ss})\approx \frac{1%
}{2}L_{y}$, it follows that the $t^{1/3}$ behavior is dominant for $\gamma
(T,L_{y},L_{x})\equiv \tau _{cross}/\tau _{ss}\ll 1$. It also
follows $\gamma \rightarrow 0$ for finite $T$ in the thermodynamic
limit. Consequently, our theory predicts that the $t^{1/3}$ growth is the
general one to be observed, as it is also concluded in \cite{mukamel}; 
the $t^{1/4}$ growth should only be observable in ``small'' $-$as 
defined below$-$ systems. It is to be noted that, in equilibrium, the 
Lifshitz-Slyozov-Wagner mechanism (evaporation from
small grains of high curvature and, after diffusion, condensation onto
larger grains) implies $\ell \sim t^{1/3}$, in accordance with experiments.

Let us define a longitudinal length, $\ell _{x}\sim t^{\phi _{x}};$\cite
{prl00,mukamel} one expects more rapid growth than transversely, $\phi
_{x}>1/3$. Concerning $t-$dependence, this length is only relevant
initially, until well-defined stripes form. This interesting result concerns
time dependence far from criticality. It is compatible with the possible
existence of two \textit{correlation} lengths near $T_{C}^{\infty }$ which
describe thermal fluctuations.\cite{prl01} Note also that the onset of the
multistripe state, when only $\ell _{y}(t)$ is relevant, may be defined by $%
\ell _{x}(\tau _{ms})=L_{x}$, i.e., $\tau _{ms}\sim
L_{x}^{1/\phi _{x}}$, which is on the same macroscopic time scale as $\tau _{%
cross}$ and $\tau _{ss}$.

The structure factor is, setting $k_x=0$ (the dependence on $k_{x}$ is only relevant 
at early times): 
\begin{displaymath}
S(k_{y};t)=\frac{1}{L_{x}L_{y}}\left| \sum_{x,y}n_{x,y}(t)\exp
[i%
k_{y}y]\right| ^{2}.
\end{displaymath}
This soon develops a peak at $k_{y}=k_{max}(t)$ that shifts towards
smaller wave number with increasing $t$. We define $\ell _{S}=2\pi /k_{%
max}$ as a measure of the mean stripe width. Alternatively, one may use the
first moment of $S$, or the slope of the straight portion in a plot of $\ln
[S(k_{y},t)]$ \textit{versus} $(k_{y}-k_{max})^{2}$, which is known as 
\textit{Guinier radius }. One may also monitor the number of stripes
$N_{s}$ and, averaging over all stripes in a given configuration, their
maximum width, $\ell _{max}$ and $\ell _{M}\equiv M/L_{x}$. 
After performing an extra average over more
than one thousand independent evolutions, all these quantities happen to
exhibit essentially the same dependence on $t;$ we denote $\ell (t)$ this
common behavior.

\begin{figure}[t]
\centerline{
\psfig{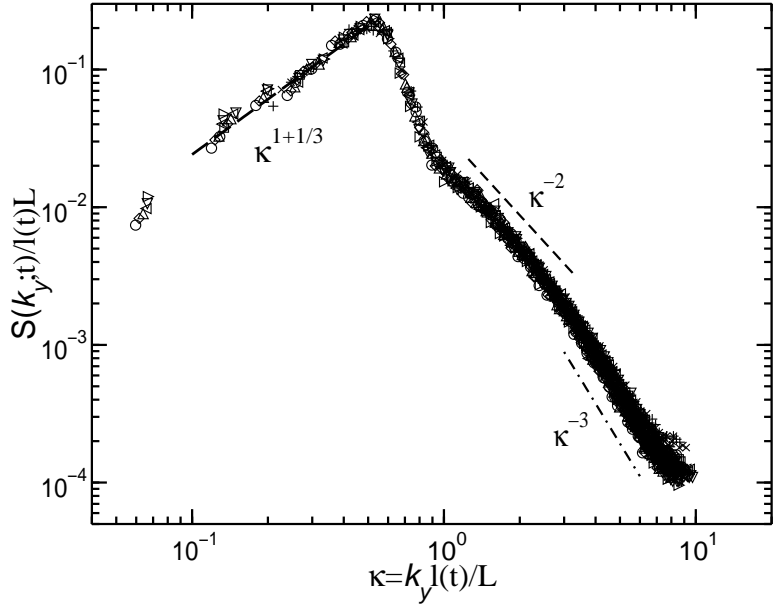}}
\caption{
Scaling with size and time of the structure
factor. The plot includes data from 64$\times $64, 128$\times $128 and 256$%
\times $256 lattices, and times $t>10^{4}$ MC steps. The
relevant behaviors (see the main text) are indicated by
dashed lines; $\kappa \equiv k_{y}\ell /L .$
}
\label{figura3}
\end{figure}

%\begin{figure}
%\centerline{
%\epsfxsize=3.5in
%\epsfbox{figura3.eps}
%}
%\caption{Scaling with size and time of the structure
%factor. The plot includes data from 64$\times $64, 128$\times $128 and 256$%
%\times $256 lattices, and times $t>10^{4}$ MC steps. The 
%relevant behaviors (see the main text) are indicated by
%dashed lines; $\varphi \equiv k_{y}\ell .$
%}
%\vspace{-0.5cm}
%\label{figura3}
%\end{figure}                                        

To check our predictions, we plotted $\ell (t)$ versus $t^{\phi _{y}}$ for
varying $\phi _{y}$ looking for the best linear fit. As illustrated in the
inset of fig. 2, $\chi ^{2}$ associated with this fit nicely
confirms that $\phi _{y}\simeq \frac{1}{4}$ for \textit{small} systems
($L_{y}<128$) while $\phi _{y}\simeq \frac{1}{3}$ for \textit{large} ones
($L_{y}\geq 256)$. This is further confirmed by studying $d \ln \ell
\ / d\ln t$ for large $\ell $ (not shown). Our theory predicts that 
this crossover with size will occur for $\tau _{cross}=\tau _{ss}$. 
For flat linear interfaces, $\langle \delta y^{2}\rangle \sim {\mathcal O}(1)$,
$\bar{\Delta}\leq 6$, $\rho _{h}\sim \exp (-16/T)$ and $\nu \simeq 1/q$,
the lattice coordination number. It then follows numerically that the crossover
from the early $\phi _{y}=\frac{1}{4}$ to the general macroscopic
$\frac{1}{3}$ behavior will occur around $L_{y}\sim 140$ for $L_{x}=64$.
Again, this is fully consistent with our observations. The reason behind is
that surfaces (and, thus, mechanism A) dominate initially, more the smaller
the system is; however, hole diffusion (B) tends to dominate as
the system relaxes towards only two surfaces enclosing the whole of the
liquid phase.

The structure factor is an important tool for experimental analysis. Given
that the DLG shows a unique $t$-dependent length after forming stripes, one
should expect $S(k_{y};t)\propto $ $\ell (t)F[k_{y}\ell (t)]$. This is
confirmed in fig. 3. (A Ginzburg-Landau model for sheared mixtures has been
shown to exhibit a similar property, though with two lengths both behaving
differently than $\ell (t)$ here.\cite{corberi}) Fig. 3 also illustrates
that data for different square lattices scale $S\ell ^{-1}L^{-1}\sim
F(\eta L^{-1}),$ $\eta \equiv k_{y}\ell$. 
A similar result does not
hold for general, rectangular lattices, as one should probably expect due to
more involved finite-size effects if $L_{x}\neq L_{y}.$

The sphericallized function for a three-dimensional system relaxing to
equilibrium was shown to satisfy $S(k,t)=J(t)\cdot F[kR(t)]$. This turned
out most useful given that $R$ and $J$ can simply be evaluated
phenomenologically as the scaling lengths for the $k$ and $S$ axes,
respectively. In particular, it then followed $F(\eta )=\Phi (\eta
)\cdot \Psi \lbrack \eta \cdot \sigma (\rho ,T)]$, with $\Phi $ and $\Psi 
$ universal functions. $\Phi $ describes the diffraction by a single grain, $%
\Psi $ is a \textit{grain interference function}, and $\sigma $
characterizes the point in the phase diagram where the sample is quenched.
In this way, it was shown \cite{prl00} that $\Psi \approx 1$ except at small
values of $k$, so that, for large $\eta $, $F(\eta )$ becomes
independent of $T,\rho $, and even the substance investigated. Though we
only obtained some weak, indirect evidence of this for the DLG (which was
investigated systematically for just one phase point), it suggests that
assuming a wide range of validity of our conclusions here is sensible.

In equilibrium, $\Phi (\eta )$ is predicted to decay according to Porod
law,\cite{rev} $\eta ^{-3}$ (for $d=2)$ at large $\eta $. In the DLG,
this concerns the region $\xi _{y}\ll k_{y}^{-1}\ll \ell (t)$, where $\xi
_{y}$ is the transverse correlation length, and the striped geometry implies
a two-point correlation function $C_{t}(x,t)\sim \frac{1}{2}(1-x\ell ^{-1})$%
, $x\ll \ell $. Therefore, we predict that $\Phi \sim \eta ^{-2}$ is the
characteristic anisotropic behavior corresponding to Porod law. This is
confirmed in fig. 3. Interesting enough, this property of the scaling
function implies $S\sim 1/\ell \left( t\right) k_{y}^{2}$ for $k_y$ 
larger than for the Guinier gaussian region described above. Finally, we
remark that the behavior $\Phi \sim \eta ^{-2}$ is
general except (for \textit{large} systems) at large enough values of $%
\eta $ (fig. 3). In this case, the standard Porod law holds, which is
induced by the very small, standard thermal clustering (which is here
similar to the one in equilibrium).

Summing up, we showed time self-similarity and simple finite-size scaling of
the structure factor in a (nonequilibrium) particle model under a large
drive following a quench to low $T$. In spite of some formal similarities,
both the scaling function and the structure factor differ from the ones for
the equilibrium (not driven) case. This is a consequence of the singular
geometry (flat linear interfaces) induced by the drive. 
The scaling function $\Phi(\eta)$ is seen to tend to an envelope 
$\sim \eta ^{1+1/3}$ at small $\eta$; it then follows a gaussian peak 
$\sim exp[-const \cdot (\eta - \eta_{max})^2]$, and decays 
$\eta ^{-2}$ and, finally, $\eta ^{-3}$ at large $\eta$.
On the other hand, there is only one $t-
$\textit{dependent} relevant length, the mean interfaces distance. This
generally grows as $t^{1/3}$ due to hole diffusion in the bulk, though one
may also observe $t^{1/4}$ at very early times due to surface
evaporation-condensation processes. The basic mechanisms are here much
simpler than in equilibrium, again due to the underlying anisotropy. That
is, the standard lattice gas typically develops grains of different sizes,
showing a distribution from microscopic to macroscopic grains. This is the
case for practically any $t$ and, as a consequence, a detailed
surface-tension-controlled kinetic description is rather complex.\cite{penr}
The DLG grain distribution, after a short transient 
time, exhibits a gap (at the, say, mesoscopic scale) between 
monomers and \textit{infinite} grains, namely, stripes of
varying width. Some of the DLG properties described here might hold in
several other anisotropic situations. We hope related experiments will be
performed that will motivate development of more complete theories.

%%%%%%%%%%%  ACKNOWLEDGMENTS %%%%%%%%%%%%%%%%%%%%%%%%%%%%
{\it Acknowledgments --} We acknowledge useful discussions with P.L. Garrido,
R. Monetti, M.A. Mu\~{n}oz, L. Rubio and G. Saracco,
and support from MCYT, projects PB97-0842 and BFM2001-2841.
%%%%%%%%%%%%%%%%%%%%%%%%%%%%%%%%%%%%%%%%%%%%%%%%%%%%%%%%%%


\begin{thebibliography}{99}
\bibitem{fnote2}
This initial regime of nucleation is theoretically
more involved, e.g., one cannot assume flat linear interfaces (as we do
below) and surface tension may play an important role. However, nucleation
extends very shortly after the melt, thus bearing relatively less practical
importance; cf. fig 1 and its caption.
\end{thebibliography}
\end{document}